# Towards a Spiking Neural P Systems OS

Ammar Adl

(ammaradl@gmail.com)

Amr Badr

(a.badr.fci@gmail.com)

Ibrahim Farag

(i.farag@gmail.com)

Computer Science Department
Faculty of Computers and information
Cairo University

Keywords:   S*piking Neural P systems, Operating System, logic Operations, Processing Neurons.*

## Abstract

This paper is an attempt to incorporate the idea of spiking neural P systems as an early seed into the area of Operating System Design, regarding their capability to solve some classical computer science problems. It is reflecting the power of such systems to simulate well known parallel computational models, like logic gates, arithmetic operation, and sorting. In these devices, the time (when the neurons fire and/or spike) plays an essential role. For instance, the result of a computation is the time between the moments when a specified neuron spikes. Seen as number computing devices, SN P systems are shown to be computationally complete, and with such capabilities, arithmetic operations, logic, and timing, some first steps could be taken towards an OS design.

## Introduction

We introduce here only some notations and the notion of register machines, for an alphabet $V$, $V*$ denotes the set of all finite strings of symbols from $V$, the empty string is denoted by $\lambda$, and the set of all nonempty strings over $V$ is denoted by $V+$. When $V = \{a\}$ is a singleton, then we write simply $a*$ and $a+$ instead of $\{a\}*$, $\{a\}+$. The length of a string $x \in V*$ is denoted by $|x|$. The family of Turing computable sets of natural numbers is denoted by NRE and the family of semilinear sets of natural numbers is denoted by NREG.[4]

A register machine is a construct M = ($m$, $H$, $l_0$, $l_h$, $I$), where m is the number of registers, $H$ is the set of instruction labels, $l_0$ is the start label (labeling an ADD instruction), $l_h$ is the halt label, and $I$ is the set of instructions; each label from $H$ labels only one instruction from $I$. The instructions are of the following forms:
- $l_1$: (ADD($r$), $l_2$, $l_3$) (add 1 to register r and then go to one of the instructions with labels $l_2, l_3$),
- $l_1$: (SUB($r$), $l_2$, $l_3$) (if register $r$ is non-empty, subtract 1 from it and go to the instruction with Label $l_2$, otherwise go to the instruction with label $l_3$),
- $l_h$: HALT (the halt instruction). [4]

A register machine M computes a number n in the following way: start with all registers empty (i.e., storing the number zero), we apply the instruction with label $l_0$ and we proceed to apply instructions as indicated by the labels (and made possible by the contents of registers); if we reach the halt instruction, then the number *n* stored at that time in the first register is said to be computed by M. The set of all numbers computed by M is denoted by N (M).

Register machines are universal also in the accepting mode; moreover, this is true even for deterministic machines, having ADD rules of the form $l_1$ : (ADD(*r*), $l_2$, $l_3$) with $l_2 = l_3$ : after adding 1 to register *r* we pass precisely to one instruction, without any choice (in such a case, the instruction is written in the form $l_1$ : (ADD(*r*), $l_2$)). Again, without loss of generality, we may assume that in the halting configuration all registers are empty.

## Spiking neural P systems

Definition: A spiking neural P system is a tuple:
$$\Pi = (O, \sigma_1, \sigma_2, \cdots, \sigma_m, syn, in, out),$$
Where:
1. $O = \{s\}$ is the unary alphabet (s is known as a spike),
2. $\sigma_1, \sigma_2, \cdots, \sigma_m$ are neurons, of the form $\sigma_i = (n_i, R_i)$, $1 \leq i \leq m$, where:
   - $n_i \geq 0$ is the initial number of spikes contained in $\sigma_i$,
   - $R_i$ is a finite set of rules of the following two forms:
     i. $E / s^b \to s$; d, where $E$ is a regular expression over *s*, $b \geq 1$ and $d \geq 1$,
     ii. $S^e \to \lambda$; 0 where $\lambda$ is the empty word, e $\geq$ 1, and for all $E/s^b \to s$; $d$ from $R_i$
     $S^e \notin L(E)$ where $L(E)$ is the language defined by $E$,
3. $syn \subseteq \{1, 2, \cdots, m\} \times \{1, 2, \cdots, m\}$ are the set of synapses between neurons, where $i \neq j$ for all $(i, j) \in syn$,
4. *in*, *out* $\in \{\sigma_1, \sigma_2, \cdots, \sigma_m\}$ are the input and output neurons respectively.

In the same manner as in [3], spikes are introduced into the system from the environment by reading in a binary sequence (or word) $w \in \{0, 1\}^*$ via the input neuron $\sigma_1$. The sequence *w* is read from left to right one symbol at each timestep. If the read symbol is 1 then a spike enters the input neuron on that timestep.

A firing rule $r = E/s^b \to s$; d is applicable in a neuron $\sigma_i$ if there are $j \geq b$ spikes in $\sigma_i$ and $s^j \in L(E)$ where $L(E)$ is the set of words defined by the regular expression *E*. If, at time *t*, rule *r* is executed then *b* spikes are removed from the neuron, and at time $t + d - 1$ the neuron fires. When a neuron $\sigma_i$ fires a spike is sent to each neuron $\sigma_j$ for every synapse $(i, j)$ in $\Pi$. Also, the neuron $\sigma_i$ remains closed and does not receive spikes, until time $t + d - 1$ and no other rule may execute in $\sigma_i$ until time $t + d$.

This does not affect the operation as the neuron fires at time $t + d - 1$ instead of $t + d$. A forgetting rule $r' = S^e \to \lambda$; 0 is applicable in a neuron $\sigma_i$ if there are exactly *e* spikes in $\sigma_i$. If $r'$ is executed then e spikes are removed from the neuron. At each timestep *t* a rule must be applied in each neuron if there are one or more applicable rules at time *t*. Thus while the application of rules in each individual neuron is sequential the neurons operate in parallel with each other.

Note that there may be two rules of the form $E/s^b \to$ s; d, that are applicable in a single neuron at a given time. If this is the case then the next rule to execute is chosen non-deterministically. The output is the time between the first and second spike in the output neuron $\sigma_m$.

An extended spiking neural P system has more general rules of the form $E/s^b \to s^p$; *d*, where $b \geq p \geq 0$. Note if p = 0 then $E/s^b \to s^p$; *d* is a forgetting rule. An extended spiking neural P system with exhaustive use of rules [4] applies its rules as follows. If a neuron $\sigma_i$ contains k spikes and the rule $E/s^b \to s^p$; *d* is applicable, then the neuron $\sigma_i$ sends out *gp* spikes after d time steps leaving $u$ spikes in $\sigma_i$, where:

$k = bg + u$, $u < b$ and $k, g, u \in \mathbb{N}$. Thus, a synapse in a spiking neural P system with exhaustive use of rules may transmit an arbitrary number of spikes in a single timestep. In the sequel we allow the input neuron of a system with exhaustive use of rules to receive an arbitrary number of spikes in a single timestep. This is a generalization on the input allowed by Ionescu et al.

In Korec's notion of strong universality was adopted for small SN P systems. Analogously, some small SN P systems could be described as what Korec refers to as weak universality. However, it could be considered that Korec's notion of strong universality is somewhat arbitrary and we also pointed out some inconsistency in his notion of weak universality. In the sequel each spike in a spiking neural P system represents a single unit of space. The maximum number of spikes in a spiking neural P system at any given timestep during a computation is the space used by the system.

## Addition

Here is a simple SN P system that performs the addition of two natural numbers. It is composed of three neurons Figure 1, two input neurons and an addition neuron, which is also the output neuron. Both input neurons have a synapse to the addition neuron. Each input neuron receives one of the numbers to be added as a sequence of spikes, that encodes the number in binary form. As explained above, no spike in the sequence at a given time instant means 0 in the corresponding position of the binary expansion, whereas one spike means 1. The input neurons have only one rule, $a \rightarrow a$, which is used to forward the spikes to the addition neuron as soon as they arrive. The addition neuron has three rules: $a \rightarrow a$, $a^2/a \rightarrow \lambda$ and $a^3/a^2 \rightarrow a$, which are used to compute the result. Formally, the SN P system for 2-addition is defined as a structure: [5]

$\Pi_{Add} = (O, \sigma_{Input_1}, \sigma_{Input_2}, \sigma_{Add}, syn, in_1, in_2, out)$

Where:

- $O = \{a\}$;
- $\sigma_{Input_1} = (0, R_{Input_1})$, with $R_{Input_1} = \{a \rightarrow a\}$;
- $\sigma_{Input_2} = (0, R_{Input_2})$, with $R_{Input_1} = \{a \rightarrow a\}$;
- $\sigma_{Add} = (0, R_{Add})$, with
  $R_{Add} = \{a \rightarrow a, a^2/a \rightarrow \lambda, a^3/a^2 \rightarrow a\}$;
- $syn = \{(Input_1, Add), (Input_2, Add)\}$
- $in_1 = Input_1, in_2 = Input_2$;
- $out = Add$.

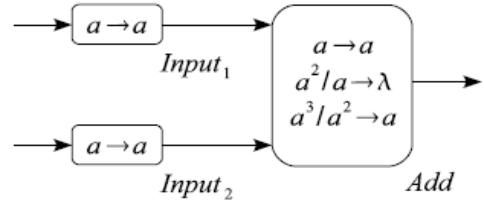

**Fig. 1.** An SN P system performs the addition among two natural numbers expressed in binary form [5]

## Checking Equality

Since an SN P system produces a spike train, we will encode the output as follows: starting from an appropriate instant of time, at each computation step the system will emit a spike if and only if the two corresponding input bits (that were inserted into the system some time steps before) are equal. Stated otherwise, if we compare two n-bit numbers then the output will also be an n-bit number: if such an output number is 0, then the input numbers are equal, otherwise they are different. [5]

Bearing in mind these marks for equality and inequality, the design of the SN P system consists of three neurons: two input neurons, having $a \rightarrow a$, as the single rule, with a third neuron, the checking neuron. This checking neuron is also the output neuron, and it has only two rules: $a^2 \rightarrow \lambda$ and $a \rightarrow a$.

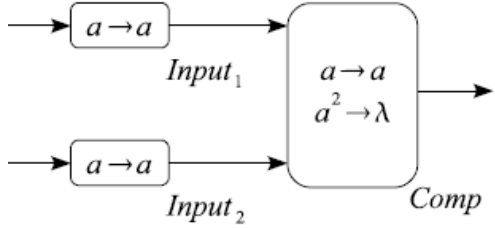

**Fig.2.** An SN P system that compares two natural numbers of any length, expressed in binary form [5].

## Simulating Logical Gates

Using SNP systems can simulate logical gates. Input is given in one neuron while the output will be collected from the output neuron of the system. Boolean value 1 is encoded in the spiking system by two spikes, hence $a^2$, while 0 is encoded as one spike. If the output neuron fires two spikes, in the second step of the computation, then the Boolean value computed by the system is 1 (hence true). If it fires only one spike, then the result is 0 (false).

Boolean AND gate simulated by SN P systems as in [6]:

$\Pi_{AND} = (\{a\},$
$\sigma_1 = (0, \{a^2 \to a; 0, a^3 \to a; 0, a^4/a^2 \to a; 0\}), \emptyset, 1).$

If in neuron 1 we introduce three spikes. This means we compute the logical AND between 1 and 0 (or 0 and 1). The only rule the system can use is $a^3 \to a; 0$ and one spike (hence the correct result - 0 in this case) is sent to the environment.

If 4 spikes are introduced in neuron 1 (the case 11), the output neuron will fire using the rule $a^4/a^2 \to a; 0$, and will send two spikes in the environment. The system with the input 00 behaves similarly to the 01 or 10 cases. If the output neuron, the rule $a^3 \to a; 0$ is changed with the rule $a^3 \to a^2; 0$, this will lead to the OR gate. [6]

For Boolean NOT gate it can be simulated by SNP systems using two neurons, no delay on the rules, in two steps [6]. With the structure:

$\Pi_{NOT} = (\{a\}, \sigma_1, \sigma_2, \{(1, 2), (2, 1)\}, 1),$

And:

$\sigma_1 = (a, \{a^2/a \to a; 0, a^3 \to a; 0\})$
$\sigma_2 = (0, \{a/a \to a; 0, a^2/a^2 \to a; 0\}).$ [6]

The initial configuration, neuron 1 contains 1 spike, which, once used to correctly simulate the gate, has to be present again in the neuron such that the system returns to its initial configuration. This is done with the help of neuron 2 which in step 2 of the computation refills neuron 1 with one spike.

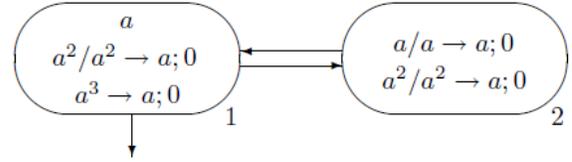

**Fig.3.** SN P systems simulating NOT gate [6]

If the input in the Boolean gate is 0, hence one spike is introduced in neuron 1, it uses the rule $a^2/a \to a; 0$, two spikes are sent to the environment (and the result of the computation is 1), and to neuron 2 in the same time. In the second step of the computation neuron 2 uses the rule $a^2/a^2 \to a; 0$ consumes the two spikes present inside, and sends one back to neuron 1. The system recovers its initial configuration.

## Sorting

Employing SN P systems can solve sorting $n$ natural numbers, not using the rules in the exhaustive way, but as in the original definition of such systems. [6] Some $n$ natural numbers encoded as spikes, one in each neuron from the first layer of the. As long as they are not empty they consume at each step a spike, and send $n$ spikes, one to each neuron from the second layer of the structure, the latter neurons have $n$ different thresholds, and have $n$ different number of synapses with the neurons from the third layer of the structure. The latter ones contain the result of the computation. As in Theorem 2 in [6];

SN P systems can sort a vector of natural numbers where each number is given as number of spikes introduced in the neural structure.

## SN P systems OS

Based upon the above, we could introduce a very first step, into the direction of a simple operating system. The concept is that if we are capable of adding, subtracting, comparing, and sorting numbers, then we can take this view to a higher level of "Jobs".

Considering that an operating system is some environment for processing tasks, or jobs, and managing some predefined resources used to serving those jobs in execution, if we employ the above devices into this direction, mapping a "job" to a number, then we can find the following:

An SNPOS is a construct of:

$SNPOS =$
$(E, D \{O, \sigma_1, \sigma_2, \cdots, \sigma_m, syn, in, out\}, J_1, J_2 \cdots, J_m)$

Where:

1. E is the operating system environment- can be represented as a P System Skin Membrane
2. D is a set of SN P systems devices, of the form: $\{O, \sigma_1, \sigma_2, \cdots, \sigma_m, syn, in, out\}$,
3. $J_1, J_2, \cdots, J_m$ are the jobs contained in environment;

    A Job is of the form: $J_i = (r_i, res_i, sc_i)$,

    - $r_i \geq 0$ is the initial number of rules (instructions) contained in $J_i$,
    - $res_i$ is a finite set of job resources.
    - $sc_i$ is the scope of job execution.

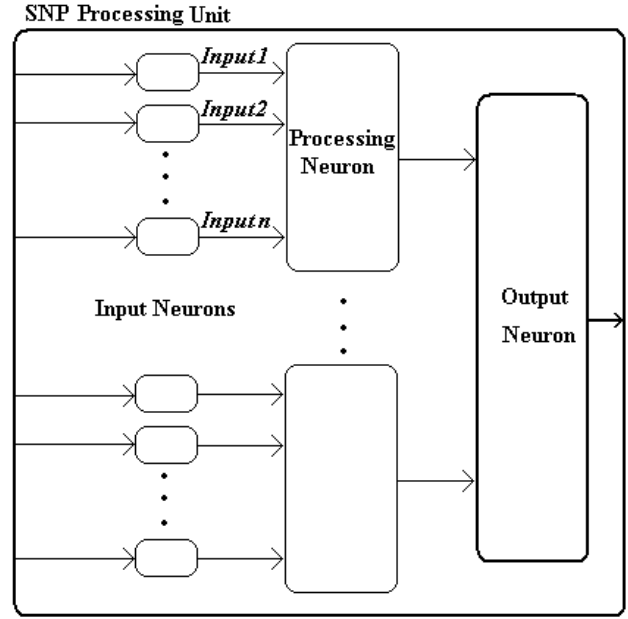

**Fig.4.** SNP Processing Unit

Given such a perspective for a SNP processing unit, we may have a bird's eye view on the OS environment building blocks as in Figure 5.

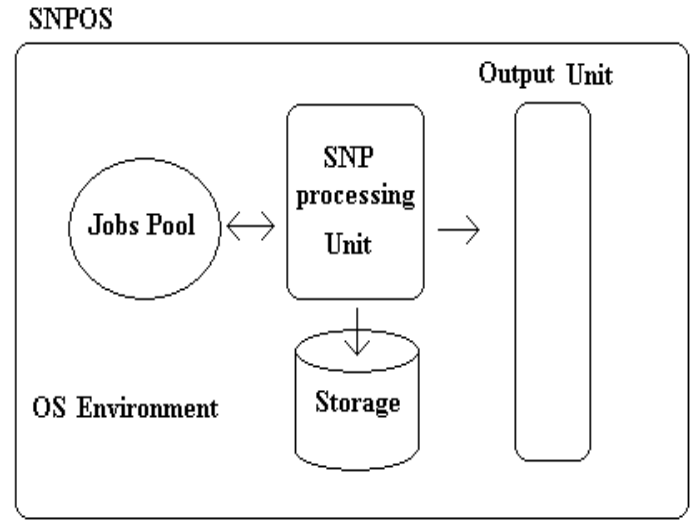

**Fig.5.** SNPOS

## Future work

For the future work, we will be focusing on the detailed design of the SNPOS building blocks, also simulating the resolution of a job and its results within the processing neurons, how they will manage the jobs' instructions execution, storage, and what will the OS environment provide as resources and algorithms for managing inputs and outputs processes.